\documentclass{elsart}
\newcommand{\be}{\begin{equation}}
\newcommand{\ee}{\end{equation}}
\newcommand{\bea}{\begin{eqnarray}}
\newcommand{\eea}{\end{eqnarray}}
\usepackage{algorithmic}
\usepackage{graphicx}
\usepackage{url}
\begin{document}
\begin{flushright}
Edinburgh 2002/17
\end{flushright}
\vspace{0.5cm}
\begin{frontmatter}
\title{SINGINT: Automatic numerical integration of singular integrands}
\author{Nikolas Kauer}
\ead{nkauer@ph.ed.ac.uk}
\address{School of Physics, University of Edinburgh, Edinburgh EH9 3JZ, UK}
\begin{abstract}
We explore the combination of deterministic and Monte Carlo methods to
facilitate efficient automatic numerical computation of
multidimensional integrals with singular integrands.
Two adaptive algorithms are presented that employ recursion and
are runtime and memory optimised, respectively.
SINGINT, a C implementation of the algorithms, is introduced
and its utilisation in the calculation of particle
scattering amplitudes is exemplified.
\end{abstract}
\end{frontmatter}
\newpage
\section*{Program summary}
{\em Title of program:} SINGINT \\
{\em Program obtainable from:} \url{http://www.ph.ed.ac.uk/~nkauer/singint/}\\
{\em Program requirements:} SINGINT can be built with any compatible C and Fortran
compilers.  It requires 
GNU Scientific Library\footnote{\texttt{http://sources.redhat.com/gsl/}} 1.2.
The program has been tested with
GNU Compiler Collection\footnote{\texttt{http://gcc.gnu.org}} 3.1
on Red Hat Linux 7.1.\\
{\em Programming language:} C, F77\\
{\em No. of bytes in distributed program:} 26224\\
{\em Distribution format:} gzip-compressed tar file\\

{\em Nature of physical problem:}\\
Efficient, robust and automatic numerical computation of multidimensional integrals 
with singular integrands, for example occurring in
the calculation of particle scattering amplitudes in perturbative field
theory.

{\em Method of solution:}\\
Runtime- or memory-optimised algorithms that combine 
integration rule and Monte Carlo techniques and employ recursion.

{\em Restrictions on the complexity of the problem:}\\
\texttt{singint\_rec}, the implementation of the recursive algorithm,
is suitable for integrands of up to 12 dimensions.
\texttt{singint\_cut}, the implementation of the global algorithm,
is currently restricted to two dimensions.
An extension of the algorithm to more than two dimensions is straightforward
and will be implemented in a future version.

\newpage

\section{Introduction}

While the practitioner can rely on highly efficient and robust
automatic routines for the numerical integration of 1-dimensional
integrals \cite{QUADPACK}, for multidimensional integrals the
situation is more complicated and no mature universal tool has
emerged yet.

To guide our discussion we start by recalling the key features of a
superior numerical integration routine: It should be efficient, robust
and automatic.  Efficient means that results of the desired precision
are obtained using as little resources as possible in terms of processor
time and memory.  Robust means that the routine yields accurate results
for a variety of integrands of interest, for example, not only
smooth ones, but also discontinuous or singular integrands.
Automatic implies that the routine requires minimal
information about the integrand, or, more precisely, that a ``black box''
that returns the value of the integrand for any point in the integration
volume is all that is required.  To devise one routine that excels
in all three categories is extremely challenging, but a number of complementary
approaches have been developed to satisfy more limited expectations.

Two basic approaches to numerical integration are deterministic integration 
with integration rules and Monte Carlo integration \cite{NumericalRecipes}.  
Both methods have strengths and weaknesses.  Integration rules, on 
the one hand, yield precise results with relatively few integrand evaluations, 
but they are not too robust and work best for very smooth functions.
Monte Carlo methods, on the other hand, impose few requirements on the integrand,
but are known to converge slowly.
To bridge the gap, adaptive procedures have been developed, that select and
evaluate suitable subregions separately, thus making deterministic integration
more robust and Monte Carlo integration more efficient.

As mentioned above, the use of Monte Carlo techniques is suggestive when
multidimensional, square-integrable functions with singularities need to be
integrated.  But, when fast integration or high precision are mandatory
pure Monte Carlo methods are not well suited.  To enable rapid convergence,
integration rule based approaches for singular integrands have been explored
in the literature \cite{GenzEspelidLyness}.
As a result, highly efficient integrators exist that exploit extrapolation methods
and can be used if the location of all singularities is known in analytic form.
However, such procedures are not automatic and not applicable if the location of
singularities is unknown or difficult to determine.

In this article we therefore present two {\em automatic\/} procedures for the
integration of low-dimensional singular integrands that aim to retain the
performance advantage of deterministic integration methods.
This is achieved by the combined use of integration rules and Monte Carlo sampling.
To the best of our knowledge, this combination has not previously been proposed
in the literature.

A second objective of this article is to explore the applicability of recursion to
adaptive integration.  Numerical algorithms have traditionally been implemented
in Fortran 77, which does not allow recursion.\footnote{Recursive behaviour can
and has nevertheless been implemented in Fortran 77 through simulated recursion,
which requires unintuitive code. We also note that this deficiency has been
remedied in the Fortran 90 standard.}  One might therefore conjecture that
recursive techniques have not been explored by practitioners to the desirable
extent.  To avoid being limited by such technicalities,
SINGINT is implemented in C, which allows to conveniently
program with recursion.\footnote{Our choice was also influenced by the fact
that the Fortran compiler of the widely-used GNU compiler collection does not
support recursion.}

The article is organised as follows: In Sections 2 and 3, we present two
efficient algorithms for the automatic numerical integration of singular integrands,
which combine integration rule and Monte Carlo techniques.
The first algorithm is runtime optimised, while the second
is fully recursive and has small memory requirements.  In Section 4, we
introduce the interface of our implementation and discuss important aspects
of SINGINT's use based on a short example. In Section 5, we give 
numerical examples and discuss a practical application in theoretical
particle physics.  The article closes with conclusions.

\section{Global Algorithm}
\label{global}

\subsection{Method}

The global algorithm we suggest decomposes the 
singular integrand $f(x)$ into a bounded function $b_c(x)$ and a singular rest
$s_c(x)$ by introducing a cut parameter $c > 0$:
\be
b_c(x) :=
  \left\{
  \begin{array}{lcl}
  +c & \mbox{if} & f(x) > c \\
  f(x) & \mbox{if} & -c \leq f(x) \leq c \\
  -c & \mbox{if} & f(x) < -c
  \end{array}
  \right.
\ee
\be
  s_c(x) := f(x) - b_c(x)
\ee

The bounded component $b_c(x)$ can then be integrated relying on automatic
deterministic methods for maximum efficiency.  To integrate the singular
component $s_c(x)$ we revert to Monte Carlo integration, an approach that 
is well suited for irregular or singular integrands and requires no
analytic knowledge about the structure of the integrand, thus being
automatic, too.
These advantages are, however, offset by significantly slower convergence
relative to deterministic methods.
With regard to our integrand decomposition, this interplay suggests the
existence of an optimal range for the cut parameter $c$.
Below that range, unnecessarily large, non-singular
regions are Monte Carlo integrated, causing slow overall convergence.
Above that range, the deterministic routine has to integrate exceedingly steep peaks
in the singular regions, necessitating a high number of function evaluations,
which again leads to slow overall convergence.  For cut parameters in the optimal
range, however, the two effects balance and efficiency is maximal.
After both components have been computed, the final result is obtained by adding
up the integrals over $b_c(x)$ and $s_c(x)$.  It is obviously independent of the
cut parameter $c$.

\subsection{Optimisation}

We employ an optimised approach to the Monte Carlo integration of $s_c(x)$
that requires a 2-dimensional grid covering the integration region.  During
the integration of $b_c(x)$, some evaluations of $f(x)$ may return values above
or below the cuts, i.e. $s_c(x) \neq 0$.  In this case the grid cell that contains
$x$ is saved.  In a second step the integral over $s_c(x)$ is calculated by using
all cells with $s_c(x) \neq 0$ that were detected in the previous step as seed cells.
Note that the number of seed cells found will typically be much smaller than the
total number of cells as long as the cut parameter $c$ is chosen 
sufficiently large.
$s_c(x)$ is integrated by using crude Monte Carlo integration to evaluate all
contributing cells using a recursive algorithm that starts with seed cells.
For Monte Carlo integration in low dimensions (less than ca. 15 dimensions
according to the analysis in Ref.~\cite{quasi-efficiency}) 
convergence can be improved significantly by using quasi-random numbers
rather than pseudo-random numbers.  We therefore base the sampling in our
implementation on the low-discrepancy sequence of Ref.~\cite{Niederreiter2}.
If the result of a particular cell is finite, all neighbouring cells are
also scheduled for evaluation.  This procedure
is applied recursively until the region where $s_c(x)$ is nonvanishing is covered.
The recursive algorithm we employ is given below as Algorithm 1 (using orientations to
identify neighbouring cells).  Note that other seed cells encountered during
execution are removed from the list of remaining seed cells.
The recursion stops when a cell evaluates to zero and is therefore at the boundary of
the integration region for $s_c(x)$.  We refer to such cells as ``border'' cells.

\begin{algorithm}
\caption{recursive coverage of convex grid area}
\begin{algorithmic}
\IF{valid cell}
   \STATE process cell
   \IF{cell is start cell}
      \STATE process cells N, NW, W, SW, S, SE, E, NE
   \ELSE
      \IF{cell type is N}
         \STATE process cell N
      \ELSIF{cell type is NW}
         \STATE process cells N, NW, W
      \ELSIF{cell type is W}
         \STATE process cell W
      \ELSIF{cell type is SW}
         \STATE process cells W, SW, S
      \ELSIF{cell type is S}
         \STATE process cell S
      \ELSIF{cell type is SE}
         \STATE process cells S, SE, E
      \ELSIF{cell type is E}
         \STATE process cell E
      \ELSIF{cell type is NE}
         \STATE process cell E, NE, N
      \ENDIF
   \ENDIF
\ENDIF
\end{algorithmic}
\end{algorithm}

The recursion spreads in a ``starlight-like'' fashion and is therefore only
suited to cover convex areas.  This limitation can be overcome with the
following extension: Apply the algorithm to one seed cell.  Then make every
encountered border cell a start cell and apply the algorithm again.  (The 
border cells of the secondary runs are not made start cells.)  This extension 
makes the procedure more robust, but made little or no difference in our test
runs.  To rule out repeated evaluations of the same cell, the status of each 
cell is tagged with one of the following labels: unknown, finite, zero.  Initially 
all cells are tagged ``unknown''.  Cells tagged ``finite'' have already been 
evaluated and taken into account, while cells tagged ``zero'' have already been 
identified as border cells.
Note that tagging each cell introduces a global 
element in the recursion, which is permissible since the algorithm used
to evaluate $b_c(x)$ is also global in nature (see below).
Once the enhanced algorithm has been run for all seed cells the integration of
$s_c(x)$ is complete.

To optimise the integration of $b_c(x)$ we utilize DCUHRE, a very efficient
and robust integrator for multidimensional integrals with {\em bounded} integrands
developed by Berntsen {\em et al.}  \cite{DCUHRE}.
Its adaptive algorithm applies integration rules to determine the
integral and error over non-uniform subvolumes until the desired total error is
achieved.  The algorithm is global, i.e.~results for all subvolumes are
retained, to be able to select at any stage the subvolume with the largest
error contribution for further refinement.
This approach is optimal in terms of runtime but expensive in terms of
memory.  We return to this issue in Section~\ref{recursive}.

\section{Recursive Algorithm}
\label{recursive}

Due to its global nature the approach described in Section \ref{global} requires
a potentially large
amount of memory, and the question arises if a viable ``local'' alternative with
small memory footprint can be found.  To that end, we propose a second,
fully recursive approach. Assume the integral $I_0$ over a hypercube with
volume $V_0$ is to be determined with precision $\Delta I_0$.  Starting with
volume $V_0$ the following procedure is applied recursively:
\begin{enumerate}
\item A value $I$ and error estimate $\Delta I$ for the integral in the cell of
volume $V$ is obtained by applying
   \begin{itemize}
   \item[a)] an integration rule (ca. 200 integrand evaluations are necessary for a
             degree 13 integration rule \cite{DCUHRE})
   \item[b)] basic Monte Carlo integration with the same number of function
             evaluations as in a)
   \end{itemize}
   If both results are compatible within errors, the one with the lower error
   estimate is selected, otherwise the result with the larger error is selected.
\item The tolerable error\footnote{This condition guarantees that the overall error
      is at most $\Delta I_0$.\label{error_condition}} in the cell is
      $\Delta I_{max} := \Delta I_0\sqrt{V/V_0}$.
\item If $\Delta I \leq \Delta I_{max}$, no further action is necessary.
      If $\Delta I > \Delta I_{max}$, the cell is divided into $n$ subcells of equal
      volume, and the integrals $I_i$ in the subcells are determined as in (1).
      
\item If $\Delta I_{div} < \Delta I/\sqrt{n}$ with $ \Delta I_{div} :=
      \left[\sum_{i=1}^n (\Delta I_i)^2\right]^{1/2}$, the procedure is applied
      recursively to the subcells.
\item If $\Delta I_{div} \geq \Delta I/\sqrt{n}$, further subdivision is not
      advantageous, and $I$ is Monte Carlo sampled until
      $\Delta I \leq \Delta I_{max}$.
\end{enumerate}

This procedure clearly shows that in general the algorithm that controls the
subdivision of the integration volume is conceptually distinct
from the algorithm used to integrate individual subcells.
To let several complementary integration methods compete in the
latter adds flexibility and should typically increase
efficiency.  Our experience in the context of the practical
application described in Section \ref{Application} confirms this
hypothesis.

\section{Using SINGINT}

To get an overview of SINGINT's interface, the header file \texttt{singint.h}
is a good starting point:
{\footnotesize
\begin{verbatim}
typedef void Integrand(const int* ndim, const double x[],
                       const int* numfun, double funvls[]);

typedef void Integrator(const int ndim, const double a[], const double b[],
                        Integrand* integrand, const double desired_abserr,
                        const double desired_relerr, const int maxpts,
                        double* result, double* abserr);

Integrator singint_mc;

Integrator singint_ir;

Integrator singint_rec;

Integrator singint_cut;
extern double singint_cut_parameter;
void singint_cut_callback(const int* ndim, const double x[],
                          const int* numfun, double funvls[]);
\end{verbatim}
}

All integrand routines have to implement the function signature \texttt{Integrand},
which essentially takes a point in the integration volume specified by array
\texttt{x} with \texttt{*ndim} elements as input and returns the
integrand value at this point in array element \texttt{funvls[0]}.
In order to avoid an additional, performance reducing integrand
wrapper when applying integration rules, pointers are used to satisfy
Fortran calling conventions and the dummy argument \texttt{*numfun}
(restricted to 1) is introduced to create compatibility with DCUHRE.

The integrand routine is one of the arguments of the
\texttt{Integrator} signature, which all integrator functions implement.
A list of all arguments with type and short description follows:
\begin{description}
\item[\texttt{ndim: const int}] number of dimensions of integral
\item[\texttt{a[]: const double}] lower limits of hypercubical integration volume
\item[\texttt{b[]: const double}] upper limits of hypercubical integration volume
\item[\texttt{integrand: Integrand*}] (pointer to) integrand function
\item[\texttt{desired\_abserr: const double}]
desired absolute error of integration result
\item[\texttt{desired\_relerr: const double}]
desired relative error of integration result
\item[\texttt{maxpts: const int}]
maximum number of integrand evaluations ($\ge$ minimum given by
\texttt{IR\_MINPOINTSPERCELL})
\item[\texttt{result: double*}]
integral estimate
\item[\texttt{abserr: double*}]
estimated absolute error of \texttt{result}
\end{description}

The global algorithm (with cut parameter) and the recursive
algorithm defined above are implemented through the functions
\texttt{singint\_cut} and \texttt{singint\_rec}, respectively.
For convenience, the underlying techniques, i.e. Monte Carlo and
integration rule-based integration, are implemented in
integrators \texttt{singint\_mc} and \texttt{singint\_ir},
respectively.  Note that \texttt{singint\_ir} is currently a wrapper
function for DCUHRE and hence per se not suited for
singular integrands (see Section \ref{global}).
A concise but complete example that illustrates how to use SINGINT
follows:

{\footnotesize
\begin{verbatim}
#include <math.h>
#include <stdio.h>
#include "singint.h"

int called_by_singint_cut = 0;

void integrand(const int* ndim, const double x[],
             const int* numfun, double funvls[])
{
  funvls[0] = singular_function(ndim, x);
  if (called_by_singint_cut)
    singint_cut_callback(ndim, x, numfun, funvls);
}

int main(int argc, char* argv[])
{
  const double a[2] = { 0., 0. };
  const double b[2] = { 1., 1. };
  double result, abserr;

  singint_cut_parameter = 100.;
  called_by_singint_cut = 1;
  singint_cut(2, a, b, &integrand, 0.01, 0., 10000000, &result, &abserr);
  called_by_singint_cut = 0;
  printf("singint_cut result:  %.10g +- %.2g\n", result, abserr);

  singint_rec(2, a, b, &integrand, 0.01, 0., 10000000, &result, &abserr);
  printf("singint_rec result:  %.10g +- %.2g\n", result, abserr);
  
  return 0;
}
\end{verbatim}
}

As shown, when using \texttt{singint\_cut} it is first necessary
to set the external global variable \texttt{singint\_cut\_parameter} to a suitable
positive value.  Moreover, additional post-processing is necessary after integrand
evaluations, which is facilitated by calling \texttt{singint\_cut\_callback}
in the integrand routine.  A transparent code organization to that effect
(involving the Boolean variable \texttt{called\_by\_singint\_cut}) is also shown in the
preceding example.

The argument \texttt{maxpts} has slightly different meanings for
\texttt{singint\_cut} and \texttt{singint\_rec}.  For the former integrator
it specifies the maximal number of integrand evaluations used when integrating
the bounded component, while for the latter integrator it specifies the
maximal number of integrand evaluations in the Monte Carlo integration of
any {\em terminal\/} subcell, i.e.~a subcell where further subdivision would
not accelerate the integration.\footnote{Note that imposing such an upper limit
in \texttt{singint\_rec} effects a runtime guarantee, but also invalidates the
general statement in footnote \ref{error_condition}.}

As pointed out above, \texttt{singint\_rec} due to its recursive nature has comparably
small memory requirements.  \texttt{singint\_cut}, however, implements a globally
optimising strategy, and increased precision hence requires the allocation of
a larger workspace.  The size of the workspace for the integration of the bounded
and singular integrand components is
\[
\!\!\!\!\!\!\!\!\!\!\!
\underbrace{\mbox{\texttt{DCUHRE\_WKSPC\_SIZE}}\cdot \mbox{4 bytes}}_{\int b_c(x)\, dx}
+ \underbrace{\mbox{\texttt{MAXNRSEEDCELLS}}\cdot \mbox{4 bytes}
+ \mbox{\texttt{NRBINS}}^2 \cdot \mbox{1 byte}}_{\int s_c(x)\, dx}.
\]
For a given integrand, the optimal value of the ratio
\texttt{MAXNRSEEDCELLS}/\texttt{NRBINS}$^2$ is approximately equal to the fraction of
the integration volume where $s_c(x)$ is nonvanishing.  The shape of this area
also suggests an adequate resolution and thus a suitable value for \texttt{NRBINS}.
More specifically, the default setting of \texttt{NRBINS} = 100 and
\texttt{MAXNRSEEDCELLS} = 10000 requires about 50kB.  For high precision
calculations in connection with the application described in Section \ref{Application}
we used \texttt{NRBINS} = 5000 and \texttt{MAXNRSEEDCELLS} = 500000, which
required about 25MB.  Without prior knowledge of the singularity structure
of the integrand, one would have set \texttt{MAXNRSEEDCELLS} to
$5000^2$, increasing the required memory by about 100MB.

With respect to $\int b_c(x)\, dx$, up to $10^7$ integrand evaluations
require\\ \texttt{DCUHRE\_WKSPC\_SIZE} = 615402, i.e.~
about 2.3MB, while \texttt{DCUHRE\_WKSPC\_SIZE} = 92307714 allocates 
352MB and allows for up to $1.5\cdot 10^9$ integrand evaluations with a degree 13
rule.\footnote{Note that in this case \texttt{maxpts} is chosen to be very
close to the maximum value of a signed 4-byte integer, i.e. $2^{31}-1$.
Changing data types in DCUHRE would allow to go beyond this implicit upper
limit.}  The function \texttt{singint\_ir\_checks} can be used to obtain
the necessary workspace size for other \texttt{maxpts} settings.

To conclude this section, we mention a difficulty for non-global
recursive algorithms in general and the \texttt{singint\_rec}
implementation in particular.  Due to the depth-first evaluation of
integral contributions, no reliable estimate for the total integral exists
until the algorithm terminates.  It is hence impossible to translate a
specified relative error to a desired absolute error while integrating.
Since relative errors are not additive,
it is also not possible to adapt the strategy employed to limit the
absolute error.  \texttt{singint\_rec} therefore accepts only absolute
error requests and requires that \texttt{desired\_relerr} be set to zero.

\section{Numerical examples and practical application}
\subsection{Numerical examples}

Three 2-dimensional integrands $f_i(x,y)$ in the unit hypercube
$U_2 := [0, 1]\times[0, 1]$
with corner and line singularities are used to construct example integrands
$g_i(x, y)$ in $C_2 := [-1, 1]\times[-1, 1]$ with singularities at
quasi-unknown locations inside the integration region.\footnote{These integrands
$g_i(x, y)$ can thus not be integrated with the non-automatic, deterministic
algorithms of Ref.~\protect\cite{GenzEspelidLyness}.}

To be specific, the integrands $f_i(x,y)$ are given by
\be
f_1(x,y) = \frac{1}{\sqrt{x^2+y^2}\sqrt[5]{x}\sqrt[3]{y}
\left((x - 0.5)^2 + (y-0.5)^2 + 0.01\right)} \approx 32.640\,,
\ee

\be
f_2(x,y) = \frac{\ln(x+y)\ln(x)\ln(y)e^{2x+y}}{\sqrt[9]{x+y}\sqrt[5]{x}\sqrt[7]{y}}
\approx -4.5849\,,
\ee

\be
f_3(x,y) = \frac{(\ln(x))^2 e^{x+y} \cos(20x)}{\sqrt[9]{x}(\sqrt[3]{y})^2}
\approx 4.1960\,.
\ee
The integrands $g_i(x,y)$ are then defined by
\be
g_i(x,y) = 
  \left\{
  \begin{array}{lcl}
  f_i(x, y) & \mbox{if} & (x,y) \in U_2 \\
  f_i(-x, y) & \mbox{if} & (x,y) \in [-1, 0]\times[0, 1] \\
  f_i(x, -y) & \mbox{if} & (x,y) \in [0, 1]\times[-1, 0] \\
  f_i(-x, -y) & \mbox{if} & (x,y) \in [-1, 0]\times[-1, 0] \\
  \end{array}
  \right.
\ee
with $i=1, 2, 3\,$.

Evidently,
\be
\int_{C_2} g_i(x,y)\;dx dy = 4 \int_{U_2}f_i(x,y)\;dx dy\,,\quad\ \ i=1, 2, 3\,.
\ee

The SINGINT test program output displayed in App.~A shows that the
global algorithm implemented in \texttt{singint\_cut}, as well as
the recursive algorithm implemented in \texttt{singint\_rec} can
be used to efficiently integrate the example integrands $g_i(x,y)$.
It also confirms that results obtained with \texttt{singint\_cut}
are indeed independent of the cut parameter.  The test program further
demonstrates that the integrands $f_i(x,y)$ as well as a number of
trivial, bounded test integrands are also computed
accurately.\footnote{These integrands can be integrated more efficiently
by other means \protect\cite{GenzEspelidLyness,DCUHRE}.}

\subsection{Practical application}
\label{Application}

\begin{figure}[htb]
\vspace*{0.5cm}
\begin{center}
\begin{minipage}[c]{.48\linewidth}
\flushright \includegraphics[width=5.5cm, angle=90]{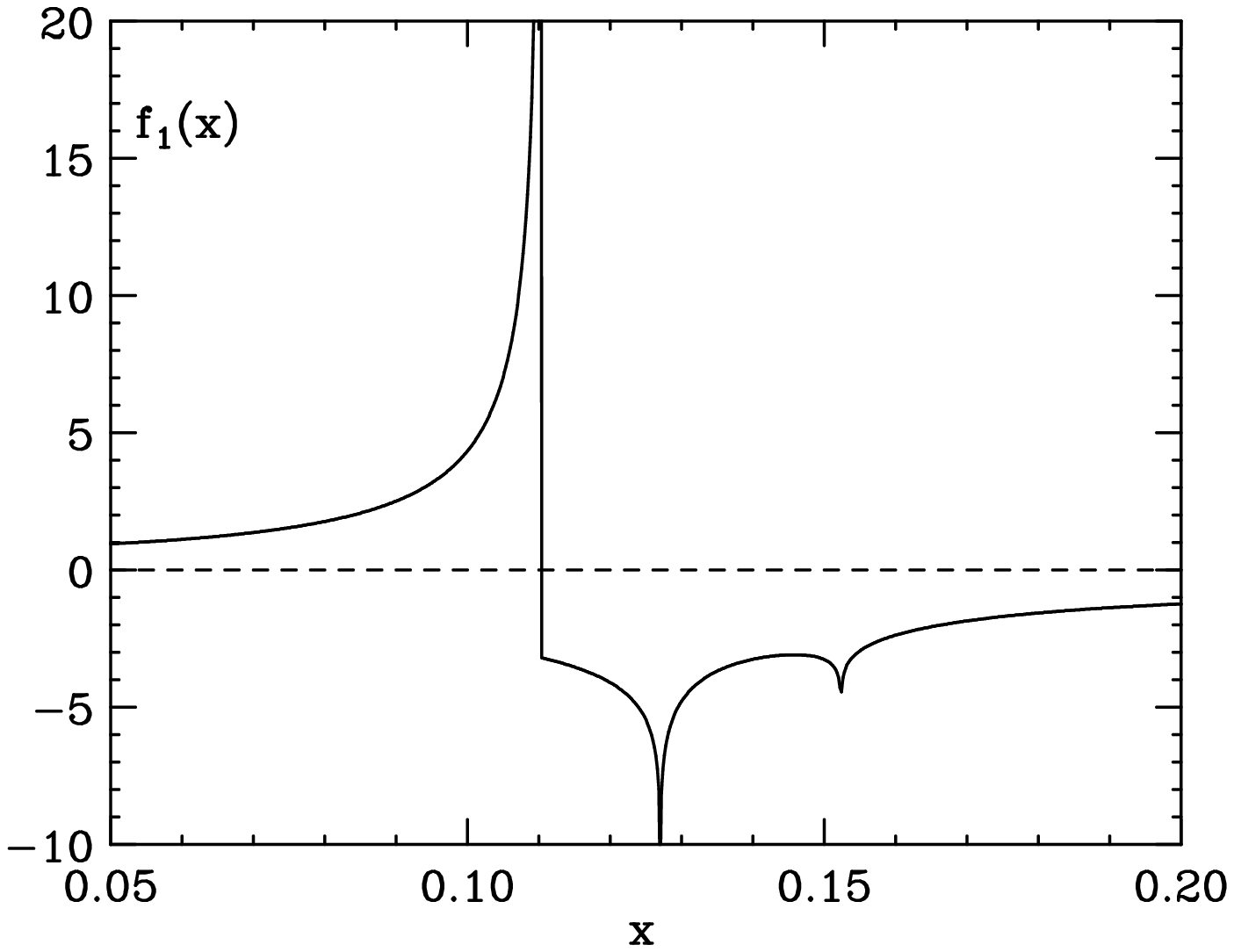}
\end{minipage} \hfill
\begin{minipage}[c]{.48\linewidth}
\flushleft \includegraphics[width=5.5cm, angle=90]{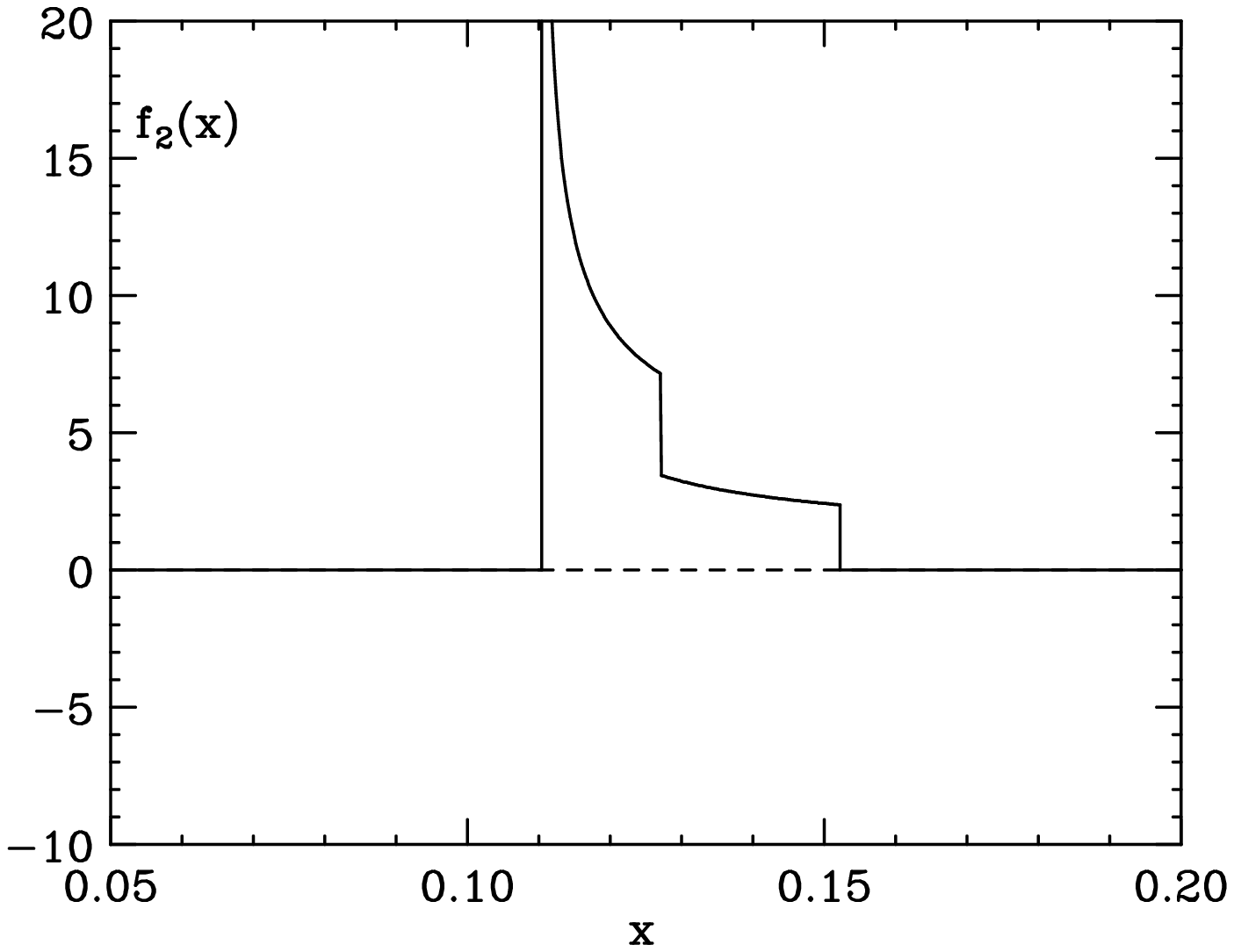}
\end{minipage}
\end{center}
\caption{Two functions $f_1(x)$ and $f_2(x)$ that illustrate the
structure of the 1- and 2-dimensional integrands that are evaluated
in the computation of multi-particle scattering amplitudes \protect\cite{Hexagon}.
The peaks correspond to integrable square-root and logarithmic singularities.
The structure of the 2-dimensional integrals is obtained by ``layering'' (with
nontrivial offset) 1-dimensional cross sections of the type shown.}
\label{singfig}
\end{figure}

SINGINT was developed to facilitate a novel numerical approach to the
calculation of multileg 1-loop amplitudes in theoretical particle
physics \cite{Hexagon}.  These amplitudes are important ingredients for
the computation of higher-order corrections to multi-particle cross
sections.
The procedure involves the evaluation of 1- and 2-dimensional
integrals that contain integrable singularities of logarithmic and
square-root type.  The number and location of these singularities
is fairly complex and depends strongly on the underlying physical
configuration as discussed in detail in Ref.~\cite{Hexagon}.
A representative example is displayed in Fig.~\ref{singfig}.

The ultimate objective is to be able to reliably compute these amplitudes for a
multitude of physical configurations in time frames of 5 to 50 seconds with
1\% relative error or better.  Initial tests with pure Monte Carlo methods
\cite{VEGAS} turned out to be 1 to 4 orders of magnitude too slow
depending on the physical configuration.  However, with the ``mixed'' algorithms
described above we were able to achieve the desired runtimes.
Generally, \texttt{singint\_cut} performed better than \texttt{singint\_rec}
for the particular type of integrand considered here.  We experimented with
cut parameter values from 500 to 50000 and found the optimal window to range
from about 5000 to 10000.
When applying the recursive evaluation of subcells
implemented in \texttt{singint\_rec} we observed that neither the
integration rule technique nor the Monte Carlo method dominated,
which confirms the advantage of a mixed strategy in the case at hand.
To further test the correctness of our implementation, we compared results 
obtained for special cases with corresponding results found in the
literature.  SINGINT also passed these application-related tests.

\section{Conclusions}

We introduced two efficient automatic procedures that facilitate
the adaptive numerical integration of low-dimensional singular
integrands.  We showed how the combined use of integration rules
and Monte Carlo sampling and the power of recursion enable new
runtime and memory optimisations.  An application in theoretical
particle physics was discussed that demonstrates the practical
value of our approach.  To allow others to apply the
algorithms, we described a C implementation with test
examples and explained its use.

With the described physics application in mind, we formulated and
implemented the global algorithm of Section \ref{global} specifically for
2-dimensional integrals.
However, its generalisation to more than two dimensions is straightforward.
There are in fact applications in perturbative field theory, where higher-dimensional
integrands similar to those of Section \ref{Application} need to be
calculated numerically.  We plan to investigate such cases in the future.

\section*{Acknowledgements}
We would like to thank T.~Binoth, A.~Genz, B.~Lautrup and J.~Lyness for useful
discussions.

\begin{appendix}
\section{Output of test program}
{\footnotesize
\begin{verbatim}
singint_ir, simple1, unit h.c.:
1 +- 1.4e-14   (1.4e-14)
singint_mc, simple1, unit h.c.:
1 +- 0   (0)
singint_rec, simple1, unit h.c.:
1 +- 0   (0)
singint_cut, simple1, unit h.c., cut = 1.5:
1 +- 1.4e-14   (1.4e-14)
singint_cut, simple1, unit h.c., cut = 0.5:
1 +- 6.8e-15   (6.8e-15)
singint_ir, simple1, non-unit h.c.:
19.2 +- 2.6e-13   (1.4e-14)
singint_mc, simple1, non-unit h.c.:
19.2 +- 0   (0)
singint_rec, simple1, non-unit h.c.:
19.2 +- 0   (0)
singint_cut, simple1, non-unit h.c., cut = 1.5:
19.2 +- 2.6e-13   (1.4e-14)
singint_cut, simple1, non-unit h.c., cut = 0.5:
19.2 +- 1.3e-13   (6.8e-15)
singint_ir, simple2, non-unit h.c.:
26.88 +- 5.9e-14   (2.2e-15)
singint_mc, simple2, non-unit h.c.:
26.88006626 +- 0.2   (0.0074)
singint_rec, simple2, non-unit h.c.:
26.88 +- 5.9e-14   (2.2e-15)
singint_cut, simple2, non-unit h.c., cut = 7.0:
26.88 +- 5.9e-14   (2.2e-15)
singint_cut, simple2, non-unit h.c., cut = 1.0:
26.88387169 +- 0.19   (0.0069)
singint_cut, singcomp1, [-1, 1]^2, cut = 50000:
129.2232525 +- 1.4   (0.011)
singint_cut, singcomp1, [-1, 1]^2, cut = 10000:
129.7198049 +- 1.4   (0.011)
singint_cut, singcomp1, [-1, 1]^2, cut = 2000:
130.1922621 +- 1.3   (0.01)
singint_cut, singcomp2, [-1, 1]^2, cut = 50:
-18.32923723 +- 0.19   (0.01)
singint_cut, singcomp2, [-1, 1]^2, cut = 10:
-18.33036408 +- 0.18   (0.0099)
singint_cut, singcomp2, [-1, 1]^2, cut = 2:
-18.3208958 +- 0.16   (0.0088)
singint_cut, singcomp3, [-1, 1]^2, cut = 50000:
16.5502603 +- 0.17   (0.01)
singint_cut, singcomp3, [-1, 1]^2, cut = 10000:
16.72847529 +- 0.17   (0.01)
singint_cut, singcomp3, [-1, 1]^2, cut = 2000:
16.79531051 +- 0.17   (0.01)
singint_rec, singcomp1, [-1, 1]^2:
129.9953115 +- 1.3   (0.01)
singint_rec, singcomp2, [-1, 1]^2:
-18.33534209 +- 0.18   (0.0098)
singint_rec, singcomp3, [-1, 1]^2:
16.7794245 +- 0.17   (0.01)
\end{verbatim}
}
\end{appendix}

\begin{thebibliography}{10}
\expandafter\ifx\csname url\endcsname\relax
  \def\url#1{\texttt{#1}}\fi
\expandafter\ifx\csname urlprefix\endcsname\relax\def\urlprefix{URL }\fi

\bibitem{QUADPACK}
R.~Piessens et al., QUADPACK, Springer Verlag, 1983.
\bibitem{NumericalRecipes}
W.H.~Press et al., Numerical Recipes, 2nd ed., Cambridge University Press, 
1992.
\bibitem{GenzEspelidLyness}
T.O.~Espelid and A.~Genz, Numerical Algorithms 8 (1994) 201 and references 
therein;
K.~Singstad and T.O.~Espelid, J.~Comp.~Appl.~Math. 112 (1999) 291;
J.N.~Lyness, Math.~Comp. 30 (1976) 1.
\bibitem{quasi-efficiency}
F.~James, J.~Hoogland and R.~Kleiss, Comp.~Phys.~Comm. 99 (1997) 180.
\bibitem{Niederreiter2}
P.~Bratley, B.L.~Fox and H.~Niederreiter, ACM Trans.~Model.~Comp.~Sim. 2 (1992) 195.
\bibitem{DCUHRE}
J.~Berntsen, T.O.~Espelid and A.~Genz, ACM Trans.~Math.~Softw. 17 (1991) 
437;
J.~Berntsen, T.O.~Espelid and A.~Genz, ACM Trans.~Math.~Softw. 17 (1991) 
452.
\bibitem{Hexagon}
T.~Binoth, G.~Heinrich and N.~Kauer, preprint Edinburgh 2002/16 (2002)
[arXiv:hep-ph/0210023].
\bibitem{VEGAS}
G.P.~Lepage, J.~Comput.~Phys. 27 (1978) 192;
G.P.~Lepage, preprint CLNS-80/447 (1980).
\end{thebibliography}
\end{document}